\documentclass{epl}

\title{Cooperative motion and growing length scales in supercooled confined liquids}
\shorttitle{Cooperative motion in supercooled liquids}
\author{P. Scheidler\inst{1} \and W. Kob\inst{2} \and K. Binder\inst{1}}
\institute{
  \inst{1} Institut f\"ur Physik, Johannes Gutenberg-Universit\"at, 
  55099 Mainz, Germany \\ 
  \inst{2} Laboratoire des Verres, Universit\'e Montpellier II, 
  34000 Montpellier, France 
}

\begin{document}

\maketitle

\begin{abstract}
Using molecular dynamics simulations we investigate the relaxation
dynamics of a supercooled liquid close to a rough as well as close to
a smooth wall. For the former situation the relaxation times increase
strongly with decreasing distance from the wall whereas in the second case
they strongly decrease. We use this dependence to extract various dynamical length
scales and show that they grow with decreasing temperature. By calculating
the frequency dependent average susceptibility of such confined systems
we show that the experimental interpretation of such data is very difficult.

\end{abstract}


\vspace*{-5mm}

\section{Motivation}
The details of the mechanism giving rise to the dramatic slowing down
of the dynamics of glass-forming liquids upon supercooling are still
unknown (see, {\it e.g.}~\cite{idmrcs4}). Although the mode-coupling
theory of the glass transition allows to rationalize many features of the
relaxation dynamics of these systems~\cite{mct}, the answers to certain
important questions (e.g. the relaxation dynamics at low temperatures)
are still unknown. A further popular approach is the {\it phenomenological}
concept of ``cooperativity'', introduced by Kauzmann~\cite{kauzmann48}, and
Adam and Gibbs~\cite{adam65}.  A typical example of this cooperativity is
the so-called ``cage-effect'', i.e. the fact that in a dense liquid each
particle is surrounded by neighboring particles which form a temporary
cage. In order to allow the particle to change its position the cage
has to open up. However, each of the particles of the cage is itself
also caged and hence can move only if other particles make room. Therefore
one can conclude that the particle motion is collective and there 
exist ``cooperatively rearranging regions'' (CRR's) within the liquid. 
The typical size of a CRR is
postulated to grow with decreasing temperature, hence ``rationalizing''
the slowing down of the dynamics~\cite{adam65,huth00}.

Experimentally it is difficult to test the concept of the CRR's since usually one
does not have direct access to the dynamics of single particles. Therefore
many studies have focused on investigating systems in spatial
confinement. If CRR's do exist and grow with decreasing temperature
the dynamics should differ from the bulk behavior as soon as the size
of the CRR's at a given temperature becomes comparable to the system
size. Indeed, almost all experiments on glass formers
confined to porous host material~\cite{pissis94,barut98,schueller94,
arndt97,jackson96,richert96,zorn02} and supported (or even free standing)
films~\cite{keddie94,wallace95,forrest97,forrest00,fukao00} do indeed show
a relaxation dynamics that differs from the one in the bulk. However,
so far it has not been possible to give a conclusive interpretation
of experimental results, since, e.g., one sometimes finds that the dynamics
in confined systems is faster than the one of the bulk, whereas in
other systems it is slower (see, e.g., ~\cite{schueller94,jackson96}).
The main reason for the diversity in experimental findings (see, e.g.,
the review article ~\cite{mckenna00gra}) is the influence of secondary
effects beyond the picture of CRR's: On the one hand density effects
can be expected to play a crucial role.  Since in experiments only the
{\it average} density is accessible, one can imagine a situation where the
confined liquid shows strong local density variations, {\it e.g.} density
oscillations due to layering effects, which will strongly influence
the dynamics as well. Also the interaction between the surface and
the liquid is most likely very important~\cite{wallace95,forrest97}.
If the liquid particles are sticking to the wall, their mobility is
strongly suppressed and because of cooperativity also the dynamics of
particles a certain distance away from the wall will be slowed down, which
in turn will influence the dynamic properties of the whole sample. The
opposite should hold in a situation where the liquid particles can slip
along the boundary. 

The goal of the present paper is to use molecular dynamics computer
simulations to investigate how these boundary effects influence the
relaxation dynamics of confined glass-forming liquids and to what extend
growing length scales can be extracted in such systems.


\section{Simulation}
The model liquid under investigation is a binary mixture
of particles inter\-acting via a Lennard-Jones (LJ)
potential $V_{\alpha\beta}(r)=4\epsilon_{\alpha\beta}
[(\sigma_{\alpha\beta}/r)^{12}- (\sigma_{\alpha\beta}
/r)^{6}]$ with $\alpha,\beta \in\{\rm A,B\}$, cut-off radii
$r_{\alpha,\beta}^c$$=$$2.5\cdot \sigma_{\alpha\beta}$ and  interaction
parameter $\epsilon_{\rm AA}=1.0$, $\sigma_{\rm AA}=1.0$, $\epsilon_{\rm
AB}=1.5$, $\sigma_{\rm AB}=0.8$, $\epsilon_{\rm BB}=0.5$, and
$\sigma_{\rm BB}=0.88$. In the following we will use $\sigma_{\rm AA}$
and $\epsilon_{\rm AA}$ as units of length and energy, respectively,
setting Boltzmann's constant $k_B=1$, and measure time in units of
$\sqrt{m\sigma_{\rm AA}^2/48\epsilon_{\rm AA}}$, where $m$ is the mass
of the particles. Previous simulation in the bulk have shown that at
low $T$ the dynamics of this mixture quickly slows down and that its
mode-coupling temperature $T_c$ is around 0.435~\cite{kob95}.

We considered two types of walls: a rough and a smooth one. The rough
wall was realized by freezing a slice of thickness $2.5\sigma_{\rm
AA}$ of the LJ liquid and applying the same LJ interaction plus an
additional hard core potential to prevent the liquid particles from
penetrating into the wall. More details on this type of wall can be
found in ~\cite{scheidler_ANDALO,scheidler_A}

To mimic a smooth wall at a location $z_{\rm W}$ we applied an external
potential of the form  $V_{\alpha \rm W} (z)= (4/45) \pi \rho_{\rm W}
\sigma_{\rm AB}^3 \varepsilon_{\alpha {\rm W}} (\sigma_{\rm AB}/(z -
z_{\rm W}) )^9$ (with $\varepsilon_{\rm AW}=1.0$, $\varepsilon_{\rm
BW}=3.0$). By choosing walls at $z_{\rm W}=-0.65$ and $z_{\rm W}=15.65$
we make sure that a film width of $D=15.0$ is realized.

The data for the rough wall presented here comes from a simulation of
a film with area $L$x$L$ and thickness $D$ ($L=12.88$, $D=15.0$), where
periodic boundary conditions are applied in the film plane. It contains
$2400$ A and $600$ B particles giving an average density of 1.2 as in
the bulk simulations in Ref.~\cite{kob95}. For the smooth wall the area
of the film was four times larger. In order to improve the statistics
we averaged the results over 16 independent samples. The equations of
motion were integrated with the velocity form of the Verlet algorithm,
using at low $T$ a time step of 0.02, and the starting configurations
for the microcanonical production runs were carefully equilibrated.

As already mentioned in the previous section it is most important to avoid
that the confining walls change the structural properties of the enclosed
fluid. This request is by construction fulfilled trivially in the case
of the rough wall. For the smooth wall, however, we found that close to
the wall strong layering effects occur~\cite{scheidler_A}. To avoid this
problem we modified the potential energy of the system by adding a term 
that coupled directly to the deviation from a constant density profile, 
i.e. configurations whose profile is not constant (within
a certain fluctuation) are energetically disfavored. This modification
does indeed allow to obtain a system with an essentially constant density
profile and we have checked that its structure is indeed the same as
the bulk one~\cite{scheidler_A}).


\section{Results}

We now discuss the relaxation dynamics of the system as a function of
temperature and distance from the wall. Although in the following we
focus on the self part of the intermediate scattering function, all the
conclusions are also valid for other observables, such as mean squared
displacements or the van Hove correlation functions~\cite{scheidler_A}.

From earlier investigations of systems confined between rough
surfaces~\cite{scheidler_EPL,scheidler_ANDALO} we know that the
local particle dynamics depends on the distance $z$ from the wall.
Therefore we introduce a generalization of the incoherent intermediate
scattering function by $F_{\rm s}({\bf q},z,t) = N_{\alpha}^{-1}
\sum_{j=1}^{N_{\alpha}}$ $\left\langle \exp \left[ i{\bf q} \cdot
({\bf r}_j(t)-{\bf r}_j(0)) \right] \delta(z_j(0)-z)\right\rangle$,
{\it i.e.} $F_{\rm s}({\bf q},z,t)$ considers only particles that at
$t=0$ had a distance $z$ from the wall~\cite{scheidler_ANDALO}. We
will only present data for A particles and wave vectors ${\bf q}$
parallel to the wall with $|{\bf q}|=7.2$, the location of the
maximum in the structure factor. Data for other values of $|{\bf q}|$
or the B particles looks qualitatively similar. At low temperatures
$F_{\rm s}({\bf q},z,t)$ shows a two step relaxation.  Hence we can
characterize the $\alpha-$relaxation time $\tau_q(z)$ by $F_{\rm s}({\bf
q},z,\tau_q)=e^{-1}$~\cite{scheidler_EPL,scheidler_ANDALO}.

\begin{figure}
\onefigure[width=14.5cm]{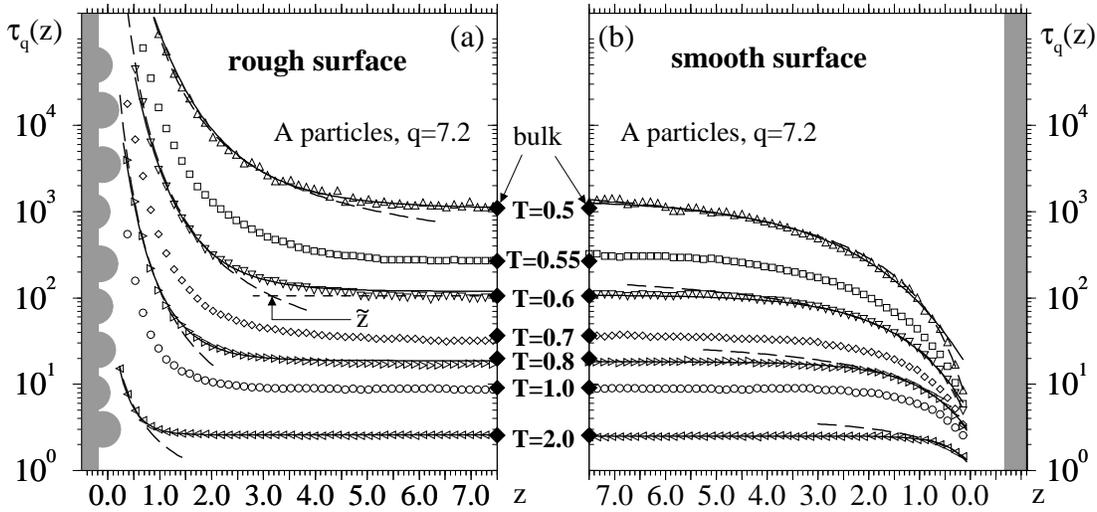}
\caption{Structural relaxation time $\tau_q(z)$ at $q=7.2$ as a function
of particle distance $z$ from the wall for (a) rough and (b) smooth
surface at different temperatures. The large diamonds are the bulk
values and the long dashed lines and the solid ones are fits according to
Eqs.(\ref{ansatz1}) and (\ref{ansatz2}), respectively.
}
\label{fig1}
\end{figure}
In Fig.~\ref{fig1} we show the relaxation times $\tau_q(z)$ as a function
of $z$ at different temperatures. For particles far away from the wall
$F_{\rm s}({\bf q},z,t)$ shows bulk behavior and therefore
the characteristic relaxation times coincide with the bulk values (filled
diamonds). Approaching the rough surface the dynamics is slowed down
dramatically, and $\tau_q(z)$ grows continuously over several decades
in time. In contrast to this the dynamics of the system with smooth
surfaces {\it accelerates} with decreasing $z$. Furthermore one can
see that, in both situations, the region affected by the wall expands
with decreasing temperature in that for high $T$ the bulk behavior is
realized already at $z \ge 2.0$, while at the lowest $T$ the influence
of the wall extends almost to the center of the film. In the following
we will use this increase to define a length scale.

In Ref.~\cite{scheidler_EPL} it was shown that close to the surface 
the $z$ dependence of $\tau_q(z)$ can be described well by the empirical Ansatz
\begin{equation}
\tau_q(z)=f_q(T) \exp \left[ \pm \Delta_q(T)/(z-z_{\rm p}) \right] ,
\label{ansatz1}
\end{equation}
with three free parameters $f_q(T)$, $z_{\rm p}=-0.5 \pm 0.15$ and
$\Delta_q(T)$, the latter weakly temperature dependent. (Here and in the
following the positive (negative) sign corresponds to systems with rough
(smooth) surfaces.)  The $z-$range for which this fit works well increases
with decreasing $T$ (dashed curves in Fig.~\ref{fig1}). The location of the
crossover to bulk behavior in the center (see data and fits for $T=0.6$
in Fig.~\ref{fig1}a) can be used to define a characteristic length scale
$\tilde{z}$ whose $T-$dependence will be discussed below.

\begin{figure}
\onefigure[width=9.5cm]{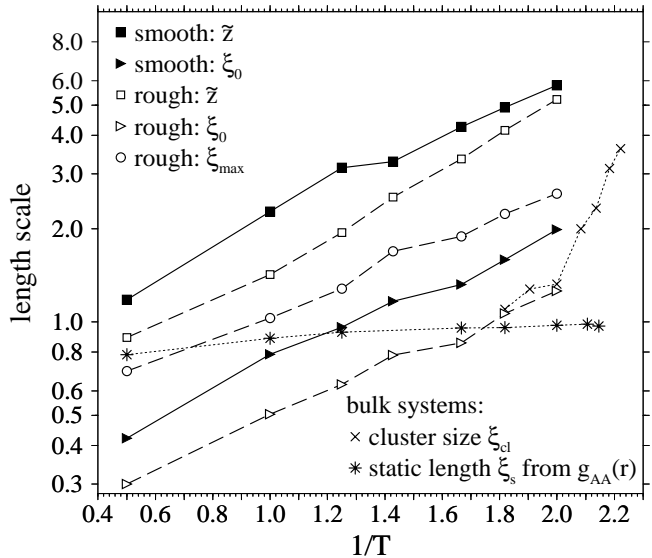}
\caption{Different dynamic length scales (see text for definition) in confined
systems as function of inverse temperature in a logarithmic plot. 
Comparison with static and dynamic length scales in bulk systems. 
Data for $\xi_{\rm cl}$ is taken from Ref.~\cite{donati98}.}
\label{fig2}
\end{figure}

An alternative Ansatz for the $z-$dependence of $\tau_q(z)$ is a function
depending on $\exp (-z/\xi_0 )$, with a characteristic length scale
$\xi_0(T)$. We find that the functional form
\begin{equation}
\ln \left[ \left( \tau_q(z) / \tau_{q,\infty} \right)^{\pm 1} \right]
= A(T) \cdot \exp \left[ - z / \xi_0(T) \right] 
\label{ansatz2}
\end{equation}
does indeed describe the $z-$dependence for all values of $z$ (solid
lines in Fig.~\ref{fig1}). Here $\tau_{q,\infty}$ is the relaxation time
of the system in the bulk.

The length scales $\tilde{z}(T)$ and $\xi_0(T)$ are obtained directly
from $\tau_q(z)$. In Ref.~\cite{scheidler_ANDALO} we have shown that for
the case of the rough surface the {\it whole} time and $z-$dependence
of $F_{\rm s}({\bf q},z,t)$ can be described very well by the Ansatz
\begin{equation}
F_{\rm s}({\bf q},z,t)=F_{\rm s}^{\rm bulk}({\bf q},t) \pm
a(t) \exp \left[- \left( z / \xi(t) \right)^{\beta(t)} \right] \, .
\label{ansatz3}
\end{equation}
The time dependence of the length $\xi(t)$ is smooth and shows a maximum,
thus showing that the influence of the wall on the dynamics is maximal
on the time scale where this maximum occurs (which is on the order of
the $\alpha-$relaxation time of the bulk). The value of the maximum in
$\xi(t)$ thus allows to define a dynamical length scale $\xi_{\rm max}$.

In the case of smooth surfaces the Ansatz (\ref{ansatz3}) is not
very useful to define a length scale. Although the data is still
described reasonably well by Eq.~(\ref{ansatz3}), $\xi(t)$ is growing
monotonically and therefore it is not possible to read off a $\xi_{\rm
max}$~\cite{scheidler_A}.

Fig.~\ref{fig2} is an Arrhenius plot of the $T-$dependence of $\tilde{z}$,
$\xi_0$ and $\xi_{\rm max}$ and from this graph we see that these
length scales grow like $\propto \exp[E/T]$. Note that the activation
energy $E\approx 1.1$ depends neither on the definition of the length
scale nor the type of wall, which gives evidence that the length scales
do indeed characterize a length scale {\it intrinsic} to the system. Within
the $T-$range investigated the growth of this scale is rather small,
{\it e.g.} $\xi_0$ grows only by a factor of $3.5$ between $T=2.0$
and $T=0.5$ while bulk relaxation times increase by several orders of
magnitude. Furthermore we find no evidence for a divergence at a {\it
finite} temperature close to the investigated temperature region, in
agreement with Ref.~\cite{scheidler_ANDALO}.

Finally we compare these length scales with the ones that can be
identified already in the bulk. Using the van Hove autocorrelation
function one can identify the most mobile particles in the
system~\cite{donati98}. It turns out that these particles are
not distributed randomly but instead form temporary clusters
(dynamical heterogenieties), the size of which grows with decreasing
temperature~\cite{donati98}.  This size is included in Fig~\ref{fig2}
as well and we find it to be comparable to the dynamic length scale from
the present simulation if $T \ge 0.5$.

As an example for a {\it static} length scale (in bulk and film systems)
we consider the decay length of $g_{\rm AA}(r)$, the radial distribution
function for ${\rm AA}$ correlations. The envelope of $g_{\rm AA}(r)-1$ is
described well by $\exp(-r/\xi_{\rm s})$, which thus defines a length
scale $\xi_{\rm s}$. This length scale shows only a very weak $T$-dependence
(Fig.\ref{fig2}), in contrast to the dynamical length scales.


\section{Interpretation and Comparison with Experiment}
Having shown how the nature of the walls influences the relaxation
dynamics of the particles of the confined system, we now discuss the
signature of the observed slowing down/acceleration of the dynamics in
an experiment in which the {\it average} dynamics of the particles
is measured. This is motivated by the fact that in a real experiment
it is usually not possible to determine the relaxation dynamics as a
function of the distance from the wall whereas the average dynamics is
directly accessible.

Thus to compare our simulations with typical experimental data we
have to calculate dynamic properties averaged over the whole system,
such as $F_{\rm s}({\bf q},t)$, the integral of $F_{\rm s}({\bf
q},z,t)$ over $z$. Although the single curves for $F_{\rm s}({\bf
q},z,t)$ at different $z$ look very similar for rough and smooth
surfaces~\cite{scheidler_EPL,scheidler_A} and the $z-$dependence
of the relaxation times $\tau_q(z)$ has the same functional form we
find a qualitative difference in the {\it averaged} curves for $F_{\rm
s}({\bf q},t)$.

\begin{figure}
\onefigure[width=14cm]{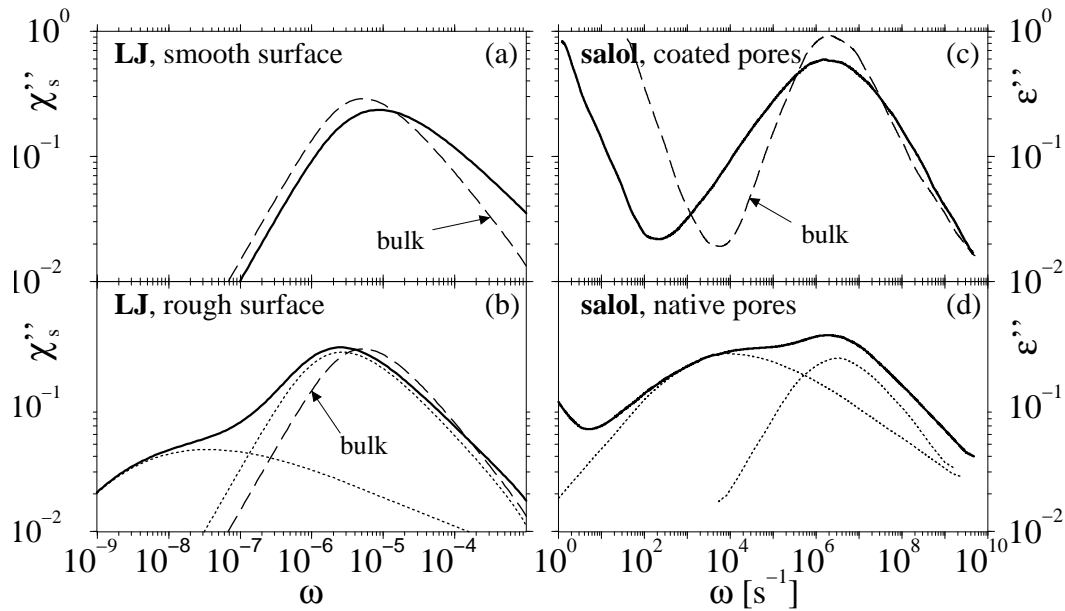}
\caption{
a/b: Dynamic susceptibility $\chi_{\rm s}''({\bf q},\omega)$ for a LJ
liquid in the bulk and confined to a film with smooth and rough surfaces
at $q=7.2$, $T=0.5$. The dotted lines are stretched exponentials. c/d:
Experimental data for the imaginary part or the dielectric susceptibility of
salol in the bulk and confined in pores. The dotted lines are fits with
Havriliak-Negami functions. Adapted from Ref.\cite{arndt97}.
}
\label{fig3}
\end{figure}

For systems with smooth surfaces the $\alpha$-relaxation is described
well by a stretched exponential law with a stretching exponent that
is slightly lower than the corresponding bulk value, {\it i.e.} due
to the superposition of different relaxation processes the curves are
more stretched.

In contrast to this, $F_{\rm s}({\bf q},t)$ for the rough walls shows a
long time tail because of the huge relaxation times for particles at the
surface~\cite{scheidler_EPL}. It is possible to describe the {\it whole}
$\alpha-$relaxation, i.e. also the mentioned tail, by the sum of two
stretched exponentials, where the time scale of the slow ``process''
is about two orders of magnitude larger than the first ``process'' and
the stretching is much more pronounced~\cite{scheidler_A}. Note however,
that this is a purely phenomenological description of the data without
any underlying physical motivation.

Typically experiments on confined liquids probe the dynamics of the
system by measuring the frequency dependence of various susceptibilities
(light-and neutron scattering scattering experiments, dielectric
measurements). Hence we have calculated the dynamic susceptibility
$\chi_{\rm s}''({\bf q},\omega) =\omega / (2 k_{\rm B} T) S_{\rm
s}({\bf q},\omega)$, where $S_{\rm s}({\bf q},\omega)$ is the time-Fourier
transform of $F_{\rm s}({\bf q},t)$.

In Fig.~\ref{fig3}a/b the frequency dependence of $\chi_{\rm
s}''({\bf q},\omega)$ at a low $T$ is shown for bulk systems as well as
for films with smooth and rough surfaces. (Note that we show only the
frequency range of the $\alpha-$ relaxation. The microscopic peak is
around $\omega=1$). Fig.~\ref{fig3}a shows that a smooth surface has a
broader $\alpha-$peak than the bulk, in agreement with the observation
that in the time domain the stretching exponent is smaller. Furthermore
the position of the peak is shifted to slightly higher frequencies, which can be
understood from the fact that close to the surface the particles move
faster than in the bulk. In contrast to these rather small differences
between the curve for the bulk and the one for the smooth surface,
the spectrum for the system with the rough wall differs strongly from
the one of the bulk (Fig.~\ref{fig3}b). First of all we note that the
location of the peak is shifted to smaller frequencies. More important
is, however, that the curve for the rough surface shows a pronounced
shoulder left to the $\alpha-$ peak. From Fig.~\ref{fig1} we know that
it would be wrong to ascribe this shoulder to the presence of a second
relaxation process, since the relaxation times are a smooth function
of $z$. Instead this shoulder is just due to the superposition of a
{\it continuum} of relaxation processes with very different relaxation
times. Also included in the figure are the Fourier-transforms of the
two stretched exponentials (dotted lines) which, noted above, describe
the whole $\alpha-$relaxation in the time domain.

The same qualitative behavior of relaxation spectra for liquids
in confinement is also seen in many experiments. As an example we
take dielectric data of the simple glass former salol confined in
Vycor glass from Ref.~\cite{arndt97}. In this ``quasi''-van der Waals
liquid H-bonds are mainly of intramolecular nature and therefore
the interaction is to a first approximation comparable to the
van der Waals system studied here. Samples with uncoated pores,
where H-bonds between the molecules and the wall can form,
correspond to the situation of a rough surface. If the pore surface is
coated, i.e. the formation of the mentioned H-bonds is prevented, the
interaction between the wall and the liquid becomes weak, {\it i.e.}
a smooth surface is realized. In Fig.~\ref{fig3}c/d we show the data
from Ref.~\cite{arndt97} for the bulk, as well as porous systems with a
native surface and a treated one. We see that from a qualitative point
of view these three spectra are very similar to the ones obtained in our
simulation for the three different situations (broadening of the peak
in the confined system, presence of a shoulder at low $\omega$ in the
untreated surface, etc.).  (Similar experimental results can be found in
Refs.~\cite{barut98,schueller94}). Also note that the increase of
the spectra at low $\omega$ is related to the Maxwell-Wagner polarization
of the sample~\cite{arndt97} which has nothing to do with the structural
relaxation of the system.) We emphasize, however, that for our system the
presence of the second peak in the susceptibility is {\it not} related to
the existence of a layer at the surface that relaxes orders of magnitudes
slower than the rest of the liquid, a popular interpretation of such a
feature~\cite{fukao00,forrest00,schueller94,barut98,arndt97,gallo00}. Instead
our analysis of the local dynamics has shown that this secondary peak
is just the result of averaging over particles that have a continuous
and {\it monotonous} distribution of relaxation times.

\acknowledgments
This work was supported by SFB 262/D1 and BI 314/18 of the DFG.
We also thank the HLRZ J\"ulich for a generous
grant of computer time on the T3E.

\end{document}